\documentclass{aastex631}

\usepackage[T1]{fontenc}

\usepackage{multirow}
\usepackage{amsmath}
\usepackage{hhline}
\usepackage{array, makecell}
\usepackage{booktabs, tabularx}
\usepackage{comment}

\newcommand{\gray}{$\gamma$-ray}
\newcommand{\grays}{$\gamma$-rays}

\usepackage{xcolor}
\newcounter{tr}
\ifnum \value{tr}>5

\newcommand{\deletedD}[1]{{\color{red} Damien - Deleted: } \sout{#1}}

\newcommand{\authorcommentD}[1]{{\color{purple} Damien - Comment :} {\color{cyan} #1}}

\else

\newcommand{\deletedD}[1]{}

\newcommand{\authorcommentD}[1]{}

\fi

\begin{document}

\title{Modeling blazar broadband emission with convolutional neural networks  - I. Synchrotron self-Compton model}

\shorttitle{Convolutional neural network for blazar SED interpretation}
\shortauthors{B\'egu\'e et al.}

\author[0000-0003-4477-1846]{D. B\'egu\'e}
\affiliation{Bar Ilan University, Ramat Gan, Israel}

\author[0000-0003-2011-2731]{N. Sahakyan}
\affiliation{ICRANet-Armenia, Marshall Baghramian Avenue 24a, Yerevan 0019, Armenia}
\affiliation{ICRANet, P.zza della Repubblica 10, 65122 Pescara, Italy}
\affiliation{ICRA, Dipartimento di Fisica, Sapienza Università di Roma, P.le Aldo Moro 5, 00185 Rome, Italy}

\author[0000-0002-8852-7530]{H. Dereli B\'egu\'e}
\affiliation{Bar Ilan University, Ramat Gan, Israel}

\author[0000-0002-2265-5003]{P. Giommi}
\affiliation{Associated to INAF, Osservatorio Astronomico di Brera, via Brera, 28, I-20121 Milano, Italy}
\affiliation{Center for Astrophysics and Space Science (CASS), New York University Abu Dhabi, PO Box 129188 Abu Dhabi, United Arab Emirates}
\affiliation{Institute for Advanced Study, Technische Universit{\"a}t M{\"u}nchen, Lichtenbergstrasse 2a, D-85748 Garching bei M\"unchen, Germany}

\author[0000-0002-0031-7759]{S. Gasparyan}
\affiliation{ICRANet-Armenia, Marshall Baghramian Avenue 24a, Yerevan 0019, Armenia}

\author[0009-0007-7798-2072]{M. Khachatryan}
\affiliation{ICRANet-Armenia, Marshall Baghramian Avenue 24a, Yerevan 0019, Armenia}

\author[0009-0007-4522-5501]{ A. Casotto}
\affiliation{Chief Scientist, Altair, 100 Mathilda Place, Suite 650 Sunnyvale, CA 94086}

\author[0000-0001-8667-0889]{A. Pe{'}er}
\affiliation{Bar Ilan University, Ramat Gan, Israel}

\begin{abstract}
Modeling the multiwavelength spectral energy distributions (SEDs) of blazars provides key insights into the underlying physical processes responsible for the emission. While SED modeling with self-consistent models is computationally demanding, it is essential for a comprehensive understanding of these astrophysical objects. We introduce a novel, efficient method for modeling the SEDs of blazars by the mean of a convolutional neural network (CNN). In this paper, we trained the CNN on a leptonic model that incorporates synchrotron and inverse Compton emissions, as well as self-consistent electron cooling and pair creation-annihilation processes. The CNN is capable of reproducing the radiative signatures of blazars with high accuracy. This approach significantly reduces computational time, thereby enabling real-time fitting to multi-wavelength datasets. As a demonstration, we used the trained CNN with MultiNest to fit the broadband SEDs of Mrk 421 and 1ES 1959+650, successfully obtaining their parameter posterior distributions. This novel framework for fitting the SEDs of blazars will be further extended to incorporate more sophisticated models based on external Compton and hadronic scenarios, allowing for multi-messenger constraints in the analysis. The models will be made publicly available via a web interface, the Markarian Multiwavelength Datacenter, to facilitate self-consistent modeling of multi-messenger data from blazar observations.
\end{abstract}


\section{Introduction}

Blazars are a subclass of active galactic nuclei which have their jet
oriented at a small angle relative to the observer's line of sight
\citep{BR78, 1995PASP..107..803U}. Due to this orientation and the relativistic
nature of their jets, blazars exhibit exceptional observational
features, such as a high luminosity, strong polarization, and rapid,
high-amplitude variability. The bolometric luminosity of blazars can reach up to
$10^{48}\:{\rm erg.s^{-1}}$ \citep[e.g.,][]{2012agn..book.....B}, making them the
most powerful non-explosive objects in the Universe.
Their extreme luminosity enables the detection of blazars even at high
redshifts \citep[e.g.,][]{RSG12,2017ApJ...837L...5A, 2020MNRAS.498.2594S, 2023MNRAS.521.1013S}.

Blazars are commonly classified into two major types
based on their optical emission lines. The blazars having
bright and broad emission lines with equivalent widths of
${\rm |EM|>}$ 5\:{\AA} are classified as
Flat Spectrum Radio Quasars (FSRQs). In contrast, when the emission lines are
weak or absent, they are identified as BL Lacertae objects (BL Lacs). While
these two subclasses share many observational similarities, the difference 
in line emission suggests that different physical mechanisms are responsible
for generating their broadband emissions. 

The broadband emission of blazars spans from radio frequencies to the high-energy
$>100$ MeV, and even to the very high-energy $>100$ GeV)
\gray\ bands, exhibiting a typical dual-bump shape
\citep[e.g.,][]{2017A&ARv..25....2P}. The low-energy component,
observed from the radio through the optical/X-ray bands,
is commonly attributed to synchrotron radiation produced by electrons
accelerated in the jet which is supported by the observed high degree of polarization \citep[e.g.,][]{2012agn..book.....B}.
However, the origin of the second component, which extends above the X-ray
band, continues to be a subject of discussion. In a leptonic
scenario, this HE component is attributed to inverse
Compton scattering of low-energy photons by the same energetic electrons
responsible for the synchrotron radiation. These low-energy photons could
either be synchrotron photons produced within the jet itself 
\citep[synchrotron self-Compton (SSC) model, see \textit{e.g.},][]{1985A&A...146..204G,
1992ApJ...397L...5M, 1996ApJ...461..657B, TMG98}, or they could have an external
origin \citep[external Compton (EC) model, see \textit{e.g.,}]
[]{1992A&A...256L..27D, 1994ApJS...90..945D, 1994ApJ...421..153S,
2000ApJ...545..107B, DFK09, GT09, SSM09}. These two alternative radiation mechanisms are also
further used to explain the differences between FSRQs and BL Lac,
respectively associated to EC and to SSC models.

Hadronic models provide another explanation of the second component: it can
either be from the direct synchrotron emission from protons that are
co-accelerated with electrons \citep{2001APh....15..121M}, or it can arise from
secondary particles generated through photo-pion and photo-pair interactions
\citep[see \textit{e.g.}][]{1993A&A...269...67M, 1989A&A...221..211M, 2001APh....15..121M, mucke2, BRS13,
2015MNRAS.447...36P, GBS22}. In this case, neutrino
emission is also expected, making blazars
attractive targets for multi-messenger astrophysical studies. The attention
to the hadronic models has grown, particularly
following the observation of IceCube-170922A, a neutrino
event which was detected from the direction of the blazar
TXS 0506+056 \citep{2018Sci...361..147I, 2018Sci...361.1378I,
2018MNRAS.480..192P}. Various models
have then been applied to explain both
the broadband spectral and the neutrino emission from individual blazars
\citep[ e.g.,][]{2018ApJ...863L..10A,2018ApJ...864...84K, 2018ApJ...865..124M, 2018ApJ...866..109S, 2019MNRAS.484.2067R,2019MNRAS.483L..12C, 2019A&A...622A.144S, 2019NatAs...3...88G, GBS22,2023MNRAS.519.1396S}.

Blazars are  monitored across various wavelengths, leading
to the accumulation of a substantial volume of multi-wavelength data over different
time periods and many numerical codes have been developed to model
this wealth of data. Some of these codes focus
exclusively on leptonic interactions. This is the case of, \textit{e.g.},
naima \citep{naima}, JetSeT \citep{TGP09,TMT11,2020ascl.soft09001T}, agnpy \citep{2022A&A...660A..18N}. Both leptonic and hadronic interactions
are included in \textit{e.g.,} AM3 \citep{GPW17}, ATHE$\nu$A \citep{MC95},
Böttcher13 \citep{BRS13} 
LeHa-Paris \citep{CZB15}, LeHaMoC \cite{SPV23} and {\it SOPRANO} \citep[Simulator of Processes in Relativistic AstroNomical Objects, ][]{GBS22}.
These codes make different assumptions, employ different methodologies,
include various physical processes, and while some operate under the steady state
assumption, others are time-dependent.

For this paper, we used the kinetic code {\it SOPRANO}. {\it SOPRANO} is a fully
conservative and implicit kinetic code designed to compute the
radiative signatures of accelerated leptons and hadrons,
taking into account a broad range of physical processes as well as
time-dependent cooling mechanisms for both primary and secondary particles.
In {\it SOPRANO}, the energy discretization is based on the discontinuous Galerking method, and the time-stepping can either be first order or exponential first order, in case of steep problems. Written in C for speed and highly optimized, {\it SOPRANO} is used via a python wrapper. This allowed us to perform the 200k simulations required for this project.
{\it SOPRANO} has been successfully applied to model the multi-messenger spectral energy distributions (SEDs) of
TXS 0506 + 056 \citep{GBS22}, PKS 0735+178 \citep[possibly in association with
several neutrino events,][]{2023MNRAS.519.1396S}, and Mrk 501 \citep{2023ApJS..266...37A}
during the historically low X-ray and \gray\ state.

Over the years, the complexity of models has dramatically increased with
the inclusion of more physical mechanisms to explain numerous observed
features and details.
For instance, including radiative contributions from protons to account
for VHE neutrinos, along with the consideration of
particle decay and cooling as they radiate, has
led to computationally intensive models, which prevent
parameter explorations and the interpretation of the data through model
fitting. As a result, fitting blazar SEDs is possible only with 'simple'
models. For example, in
\citet{2021MNRAS.504.5074S, 2022MNRAS.513.4645S, 2022MNRAS.517.2757S},
blazar SEDs observed during different periods are modeled with
JetSeT \citep{2020ascl.soft09001T}. Their analysis assumed an ad hoc electron
distribution function, and although this approach allows for estimating the
evolution of parameters over time, it does not include electron cooling. So it remains unclear whether such an ad hoc 
electron distribution  can be formed. Alternatively when computationally intensive models are built, they are
typically superimposed onto data from a specific celestial object. In
such cases, obtaining statistical information about model parameters becomes infeasible due to the prohibitive computational cost of model evaluation. 

Recent attempts to compare multi-messenger sets of data, including particle cooling and interactions have also
been made. However, among other challenges, these approaches necessitate tremendous
computational resources, questioning their use on large sample and time-resolved SED modeling. For instance, \cite{FDB08} uses a recursive strategy to attempt to converge towards the best fit parameters. A similar method, although modified, was
also used in \cite{PDP15}. Instead \citet{2017A&A...603A..31A} used a grid-scan strategy to model the SED
of Mrk 501. \citet{RPG23} also relied on a strategy of grid scanning to find the
best parameters, working in a hierarchical way from the simplest leptonic model to the most
complicated hadronic models by adding components and freezing the parameters of the previous sub-models. With this approach, no model comparison can be performed and the reliability
of the parameter distributions is impacted by the lack of cross-correlation
between the parameters at different levels, even if in the last stage a global
likelihood minimization is performed. Their study extracts parameters from 324 blazars
but requires a computational cost of approximately 17 thousand node-hours, which, to our
understanding, cannot be reused for blazars outside of the original sample.

Another recent example is the work of \cite{SPV23}, who introduced LeHaMoC, a
versatile lepto-hadronic code capable of computing spectra in just a few seconds. This
speed enabled the authors to fit the SED of the blazar HSP J095507.9+355101. However, as
acknowledged by \cite{SPV23}, the computational time required still prohibits the use of Markov
Chain Monte Carlo (MCMC) fitting for blazar SEDs. The computational time for LeHaMoC is
somewhat comparable to that of {\it SOPRANO} \citep{GBS22}, leading us to the same conclusion:
current computational resources do not permit a systematic comparison between model and data,
nor do they allow for thorough constraints on model parameters and their study.

We are therefore at a crossroad where we either continue to rely on
simple models or we find a solution that allows the use of computationally
intensive complex models for the analysis and fitting of blazar SEDs.
The objective of this paper is to introduce a new methodology that addresses
this challenge by integrating complex and resource-intensive numerical models
in detailed comparisons with data. Our method uses convolutional neural network
(hereafter CNN), a specific type of feed-forward neural network that
efficiently calculates the resulting spectrum from a given set of model
parameters with high accuracy, requiring approximately a millisecond. This
makes it well-suited for complex fitting procedures. Although the creation of the set of spectra required to train the CNN
demands considerable computational resources, once trained for a specific model,
the CNN can be cost-effectively deployed for the interpretation of
\textit{any} blazar SED.

In this paper, we train our CNN on a sample of spectra numerically obtained
from an SSC model of blazars, using {\it SOPRANO} \citep{GBS22}. We subsequently employ the trained CNNs to fit
the broadband SEDs of Mrk 421 and 1ES 1959+650 in order to demonstrate
its performance. The paper is organized as follows: In section \ref{sec:model},
we review the SSC model and outline the numerical methods implemented in {\it SOPRANO}
for computing the resulting spectra. Section \ref{sec:numerical_model} presents
our numerical table model, detailing the range of model parameters
and validating the computed spectra. Section \ref{sec:CNN} describes the CNN,
providing insights into the training procedure and the measures taken to
prevent spurious oscillations in the spectra generated by the CNN.
Section \ref{sec:example} applies the CNN to the analysis of the SEDs of blazars Mrk 421 and 1ES 1959+650 performed in the Bayesian framework. Our conclusions are summarized in section \ref{sec:conc}.

\section{The model: synchrotron self-Compton}
\label{sec:model}

\begin{table*}
\centering
\begin{tabular}{|c|c|c|c|c|c|}
    \hline
     Parameter & Units  & Symbol  & Minimum & Maximum & Type of distribution \\ \hhline{|=|=|=|=|=|=|}
     Doppler boost & -  & $\delta$ & 3 & 50 & Linear \\ 
     Blob radius        & cm &     R    &  $10^{15}$   &  $10^{18}$  &  Logarithmic \\
     \makecell{Minimum electron injection \\ Lorentz factor} & - & $\gamma_{\rm min}$ & $10^{1.5}$ &  $10^5$ &  Logarithmic \\
     \makecell{Maximum electron injection \\ Lorentz factor} & - & $\gamma_{\rm max}$ & $10^2$ &  $10^8$ & Logarithmic \\
     Injection index & - & $p$ & $1.8$ & 5 & Linear \\
     Electron luminosity & erg.s$^{-1}$ & $L_e$ & $10^{42}$ & $10^{48}$ &  Logarithmic \\
     Magnetic field  &  G  & $B$  & $10^{-3}$  & $10^{2}$ & Logarithmic  \\ \hline
\end{tabular}
    \caption{Characteristics of the dataset. For each parameter, we recall its unit (if any) and symbol, and we provide its range and the distribution of the discreet parameter values. The total number of spectra is set to $2\times 10^5$. }
    \label{tab:table_parameters}
\end{table*}

In this paper, we focus on modeling the emission from BL Lacs  within the framework of the SSC model, for which the low-energy bump is attributed to the synchrotron emission of relativistic electrons, while the second peak arises from the inverse Compton scattering of the synchrotron photons on the same electron population. This model successfully reproduces the observed multiwavelength spectrum as well as the observational features in different bands, and is widely adopted for modeling the observed data from optical to the VHE \gray\ bands.

In the one-zone SSC model, it is assumed that the emission originates
from a spherical region of the jet (referred to as a 'blob') with a
comoving radius $R$, which moves with Lorentz factor $\Gamma$. We assume that the observers sees the jet at angle $1/\Gamma$, such that the Doppler boost factor $\delta \equiv \Gamma$. The magnetic field $B$ inside this region is assumed to be homogeneous and constant. Electrons, once injected into this region, lose their energy under the effect of the magnetic field as well as by interacting with the local photon fields, ultimately generating the observed broadband spectrum.

Despite the likely presence of protons in the jet, for the SSC model, we assume that only electrons are accelerated and radiate once injected in the radiation zone. The injection function $Q_e$ is assumed to be a cutoff power-law with index $p$ for electrons with a Lorentz factor $\gamma$ larger than a minimum Lorentz factor  $\gamma_{\rm min}$, such that
\begin{equation}
Q_e  = \left \{ \begin{aligned}
& Q_{e,0} \gamma^{-p} \exp \left ( -\frac{\gamma}{\gamma_{\max}} \right ) &  &    \gamma \geq \gamma_{\rm min}, \\
&0 & & {\rm otherwise.}
\end{aligned} \right.
\end{equation}
where $\gamma_{\rm max}$ is the cutoff electron Lorentz factor. The normalization $Q_{e,0}$ is set so that the electron luminosity $L_e$ is determined by
\begin{equation}
    L_e = \pi R^2 \delta^2 m_e c^3 \int_1^{\infty} \gamma Q_e d\gamma,
\end{equation}
where $m_e$ is the electron rest mass and $c$ is the speed of light.
The temporal evolution of the electron distribution is obtained by solving the Fokker-Planck diffusion equation, while the evolution of photons is described by an integro-differential equation. We label the distribution function of photons by $N_\gamma$, and that of electrons by $N_e$. With the photon energy denoted as $x$, the kinetic equations are
\begin{align}
\left \{  \begin{aligned}
    \frac{\partial N_e}{\partial t} (\gamma ) &= \frac{N_e}{t_{\rm esc}} + \frac{\partial}{\partial \gamma } \left [  ( C_{\rm IC} N_\gamma + C_{\rm sync}  ) N_e \times \right ] + Q_{\gamma \gamma \rightarrow e^+e^- },    \\
    \frac{\partial N_\gamma}{\partial t} (x) &= \frac{N_\gamma}{t_{\rm esc}} + Q_{\rm  sync} + R_{\rm IC} N_\gamma - S_{\gamma \gamma \rightarrow e^+e^-}, 
\end{aligned} \right. \label{eq:kinetic_equation}
\end{align}
where \( t_{\rm esc} = t_{\rm dyn} = R/c \) is the escape time, such that the first term on the right hand side of each equation represent the escape of particles from the radiation zone, $C_{\rm IC}$ and $C_{\rm sync}$ represent inverse Compton and synchrotron cooling, $Q_{\gamma \gamma \rightarrow e^+e^- }$ and $S_{\gamma \gamma \rightarrow e^+e^-}$ are the source and sink terms associated to pair creation respectively, and $R_{\rm IC}$ is the redistribution kernel of Compton scattering. We note here that we do not include synchrotron self-absorption in our analysis as it is not yet included into {\it SOPRANO}. More details on the kinetic equations and their numerical solutions are given in \citet{GBS22}, which also provides the expressions for all the rates that appear in these equations.

In this paper, we employ {\it SOPRANO} \citep{GBS22} to solve the set of
coupled kinetic equations as defined in Equation \ref{eq:kinetic_equation}. 
We obtain the equilibrium solution to the kinetic equations
\eqref{eq:kinetic_equation} by evolving the system in time until $t = 4 t_{\rm dyn} = 4R/(\delta c)$. Our experiences show that further time evolution does not significantly alter the distribution functions; hence, we designate these as equilibrium distributions. These distributions serve as the final output from {\it SOPRANO} and are subsequently used to train the CNN.

\section{Numerical model: computation and validation} \label{sec:trainig_set}

\label{sec:numerical_model}

In this section, we provide details of the methodologies employed in our study to simulate spectra, which will be used as inputs to the CNN. Namely, we give details on the parameter space used for generating the SEDs via {\it SOPRANO}. With regards to the large number of spectra, we also provide our methodology to assess the validity of the generated spectra.

\subsection{Parameter ranges and sampling}

For the SSC model considered in
this paper, there are seven free parameters: the comoving blob radius $R$, the Doppler factor of the emission region $\delta$, the comoving magnetic field strength $B$ within the emission zone, the electron luminosity $L_e$, the minimum Lorentz factor $\gamma_{\text{min}}$, the cutoff Lorentz factor $\gamma_{\text{max}}$, and the power-law index $p$.
These parameters are inputs to {\it SOPRANO} which computes the resulting spectrum
in a time frame ranging from several tens of seconds to a few minutes. This
computational demand makes direct fits impossible due to the necessity to
evaluate the model tens of thousands of times for a single fit\footnote{This large number of likelihood estimations is due to the large number of parameters and is required for a full parameter exploration, for the computation of the posterior distributions and of the Bayesian factor.}. To overcome
this challenge, we developed a CNN,
which we trained on a set of $2\times 10^5$ spectra
computed by {\it SOPRANO}. The input parameters cover the whole range of
parameters relevant for an SSC model for \textit{any} blazars. The calculation of so many spectra was facilitated by coupling
{\it SOPRANO} as the spectrum generator with \textit{ronswanson}—a python-based
code designed for High-Performance Computing systems—as the distribution
software \citep{Bur23}. The code \textit{ronswanson} provides a flexible and
comprehensive interface for constructing table models from computationally
intensive simulations.

The ranges and sampling distributions for the model parameters are detailed in
Table \ref{tab:table_parameters}. The Doppler boost factor varies linearly
between 3 and 50, and the power-law index $p$ is sampled linearly within the
range of 1.8 to 5. We note that steep values of $p > 3$ are not expected from theory of shock acceleration or magnetic reconnection \citep[see e.g.][]{KGG00, SS11, Uzd22}. They are included so the range of $p$ is sufficiently large to not have to deal with boundaries. Alternatively, our method allows to set $p$ or to specify an informative prior, which can only be achieved if the model is trained on larger than expected range of the index. In contrast, the other model parameters, \textit{i.e} the
emission radius $R$, the minimum and maximum Lorentz factors $\gamma_{\rm min}$
and $\gamma_{\rm max}$, the electron luminosity $L_e$, and the strength of the
comoving magnetic field $B$, are sampled logarithmically within their respective
ranges, such that $15 < \log(R) < 18$, $ 1.5 < \log(\gamma_{\rm min}) < 5 $, $ 2 < \log(\gamma_{\rm max}) < 8 $, $ 42 < \log(L_e) < 48 $ and $ -3 < \log(B) < 2 $. This large range of the parameters guaranties that the CNN we developed will be usable for the modeling of any blazar SED.

We use Latin hypercube sampling to select the parameters of the spectra to be
computed with {\it SOPRANO} \citep[see \textit{e.g.}][]{MBC00,Via16}. This sampling
method is a widely popular technique in the creation of surrogate models as it
presents several advantages. First, it allows to specify the number of
simulations to be computed. As a byproduct, this method does not require
to specify parameter spacing. Second,
it ensures uniform sampling across all parameters. Lastly, it avoids the regular
sampling of parameters, which is typical in grid scan techniques. This
variability in the sampling enhances the performance of the CNN, see \textit{e.g.}
\cite{Kam22}.

\subsection{Properties and validation of the computed spectra}

In this section, we discuss the computational performance of {\it SOPRANO},
assessing the reliability of the computed spectra. Given that it is
impossible to individually verify each of the $2\times10^5$
computed spectra, we rely on the meta-data taken for each simulations to assess the
overall reliability of our numerical model. We anticipate that future
implementations involving more complex models of blazar SEDs, such as
external Compton or hadronic models, will necessitate even larger datasets. 
The validation methodology developed here will be applied in these future cases.
In particular, we study (i) the time to solution ensuring it aligns with
our expectations and prior experience with {\it SOPRANO}, (ii) the maximum error
of the Newton-Raphson scheme over a simulation, and (iii) the number of
times this maximum was larger than the targeted uncertainty in the computation, here set to $10^{-15}$.

\begin{figure*}
    \centering
    \includegraphics[width = 0.95\textwidth]{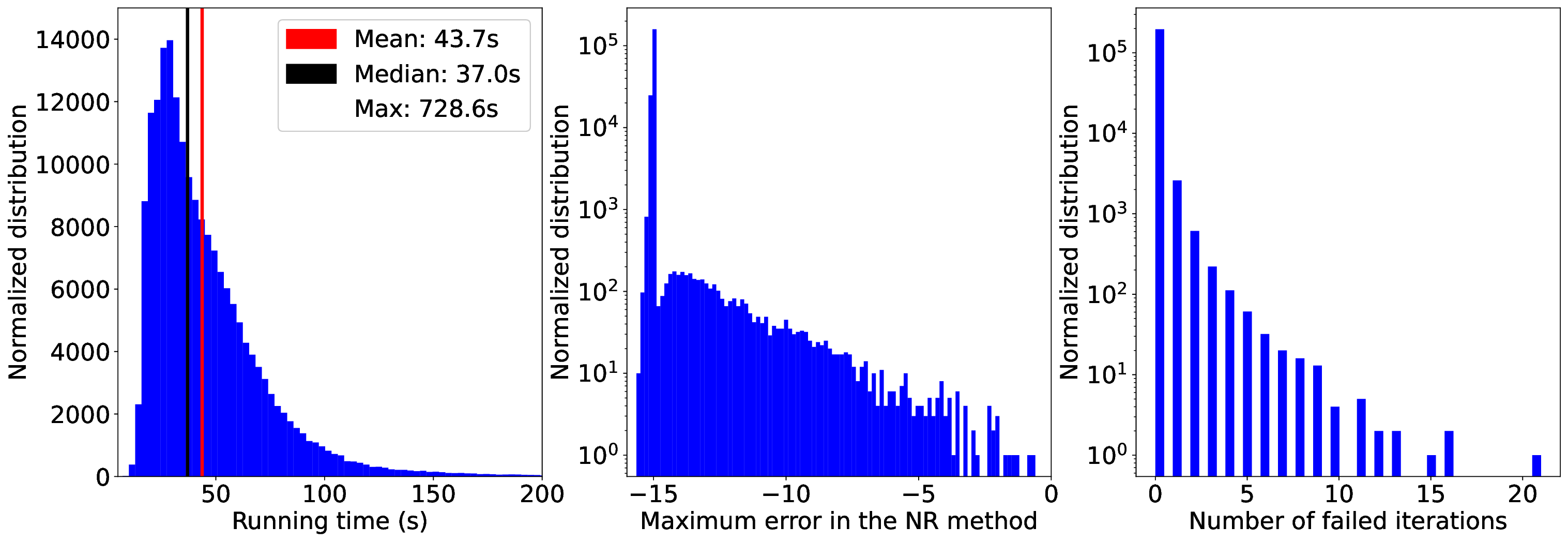}
    \caption{\textit{Left panel:} distribution of the computation time for all simulations in the table model. The average compute time is $43.7$s while the median time is $37$s. \textit{Middle panel:} Distribution of the maximum error in the Newton-Raphson method over the course of one simulation. We note that the ordinate is in logarithmic scale. Most of the spectra achieve the targeted relative error of $10^{-15}$. On the other hand, the computation of a small fraction of spectra (1.8\%) is seen to fail with a larger error. \textit{Right panel:} number of failed time iteration by simulations, which is 0 for most of our simulations, while less than 1.8 \% of our simulations have a number of failures larger than 0.  }
    \label{fig:distribution_compute_time}
\end{figure*}

First, we begin by analyzing the computational time required by {\it SOPRANO} for
each run. The left panel of Figure
\ref{fig:distribution_compute_time}
shows the the histogram of the run times for all simulations. The
average simulation time is 43.7s per spectra, with a long tail extending beyond
700 seconds. These extended durations correspond to spectra characterized by
a high compactness with small radius $R$, large electron luminosity $L_e$ and
small injection Lorentz factor $\gamma_{\rm min}$. We further note that these
computation times are obtained when each independent simulations is executed
on 8 cores on a AMD EPYC 7713 64-Core Processor CPU. An average computation time of $\sim 40$s for
evolving the spectra until 4$t_{\rm dyn}$ aligns with our initial expectations
and previous experience with {\it SOPRANO}. Overall the computation of the table model
with $2 \times 10^5$ spectra required $\sim 20$ thousands core hours, which is
feasible by any dedicated server in a couple of weeks. Although it remains a
moderately expensive computation, our approach present the advantage that it
needs to only be performed once, if the full parameter space relevant for blazar modeling in the SSC scenario is covered.

The computation of the spectra by {\it SOPRANO} can fail, specifically in regions of
large compactness, for which the numerical integrator currently used is not adapted.
These failures originate from the implicit nature of the integration scheme, which
necessitates to find the root of a non-linear systems of equations. This solution
is obtained with the Newton-Raphson root finding algorithm, which can, in some
instances, not converge towards the solution with the required accuracy. For the
current numerical model, the accuracy of the root solver is set at $10^{-15}$, close
to machine accuracy. Yet, even if the required accuracy is not reached, the photons
and electrons spectra are returned and the computation continued. Therefore, we
computed the number of failures for each spectral computation as well as the maximum
relative error on the solution.

The total number of spectra with at least one failed time iteration is 3693, 
constituting less than 2\% of all calculated spectra. The distribution of the
number of failed time bins per simulation is depicted in the right panel of
Figure \ref{fig:distribution_compute_time}. The distribution of the maximum
error across a full simulation is shown in the middle panel of Figure
\ref{fig:distribution_compute_time}. It is evident that only a small fraction
of the spectra are unreliable, with most spectra having a maximal error below
$10^{-10}$. We verified that the unreliable spectra
are in the range of parameters space which are irrelevant for the interpretation of
blazar SED.

\section{Convolutional Neural Network}

\label{sec:CNN}

We initially attempted to directly use a table model by performing multi-dimensional
linear interpolation between the intput parameters to evaluate the model for any
given parameter set. However, we encountered limitations in the interpolation
procedure in a critical region of the spectrum, specifically at the transition
between the synchrotron and SSC components. Even increasing the number of points
in the table model to several millions did not resolve this issue. This transition
frequently occurs in the X-ray band and must be accurately represented for detailed
analysis. Furthermore, the accurate modeling of this transition is also crucial in
scenarios where neutrinos could be produced, as it constrains the maximum proton
luminosity \citep[e.g., see][]{GBS22, 2023MNRAS.519.1396S, KMP18, SPG22}.

To address the challenge of fitting blazar SEDs, we have developed a surrogate model utilizing a
CNN. In essence, the CNN is modeling the relationship between input parameters
and their corresponding spectra. Our CNN is designed  to reproduce the spectra from {\it SOPRANO} in 150 energy bins. Before performing the training, the input parameters are detrended and their mean removed. We follow the same receipt for the spectra.  However, instead of considering the 150 output independent, the mean
and variance are computed for all outputs across all generated spectra. This is an
essential step because these outputs are not truly independent: they collectively
form a consistent spectrum. Based on our experience and trials, treating the
averages and means as independent variables leads to less accurate reconstructions. 
Furthermore, if each output is considered independently of the other, unwanted
oscillations appear in the produced spectra. This is because if each values are
independent, each one can overestimate or underestimate the spectrum independently
of each other. To remove these oscillations, we introduce three linear
combinations that link together the 150 spectral outputs within the model, by
constraining linear combinations of local neighbors. These combinations
are chosen to represent the finite difference derivative at order 2 and 8, as
well as the finite difference of the second order derivative at order 4. In other
words, our output vector is of length 586 where
\begin{itemize}
    \item the first 150 outputs represent the targeted spectral output,
    \item the next 142 outputs represent the $8^{\rm th}$ order finite difference of the first derivative, multiplied by a numerical coefficient $d_8^1$ namely 
    \begin{align}
        \frac{df_i}{d\epsilon} &= d_8^1 \left [ \frac{f_{i-4}}{280} - \frac{4f_{i-3}}{105} +  \frac{f_{i-2}}{5} -  \frac{4f_{i-1}}{5}  \right. \\
         & \left. ~~~~~~~ ~~~ + \frac{4f_{i+1}}{5} - \frac{f_{i+2}}{5} + \frac{4f_{i+3}}{105} - \frac{ f_{i+4}}{280} \right ]
        \nonumber
    \end{align}
        \item the next 148 outputs represent the 2nd order finite difference approximation of the first derivative, multiplied by the coefficient $d_2^1$, namely
    \begin{align}
        \frac{df_i}{d\epsilon} = d_1^1 \left [-f_{i-1} + f_{i-1} \right ]
    \end{align}
        \item the last 146 outputs represent the 4th order finite difference approximation of the second derivative, multiplied by the coefficient $d_4^2$ namely
    \begin{align}
        \frac{df_i}{d\epsilon} =  d_4^2 \left [ - \frac{f_{i-2}}{12} +  \frac{4f_{i-1}}{3} -  \frac{5f_{i}}{2} +  \frac{4f_{i+1}}{3} - \frac{f_{i+2}}{12} \right ]
    \end{align}
\end{itemize}
The CNN computes the 150 initial spectral outputs, and the remaining linear
combinations are added in a last linear step. We find that setting the
normalisation coefficients to $d_4^1 =  10$, $d_2^1 = 2$, and $d_4^2 = 4$ provides
an adequate balance between (i) learning rate, (ii) accuracy of the CNN and (iii)
the smoothness of the solution, specifically characterized by the absence of oscillatory behavior in the output spectra. We actually found that this method also increases the learning rate and the accuracy of the CNN.

By recursively building the CNN, we have determined that
a deep network is not necessary to produce an accurate representation of our
numerical model, which is computed using {\it SOPRANO}. Indeed, our CNN contains only 8
layers in this order: a first dense layer transform the 7 inputs to a high dimensional vector, 5 1D convolutional layer with different kernel size and stride, one maxpooling layer followed by a 1D convolutional layer and a final dense layer, mapping to the 150 outputs. This final layer of length 150 is then multiplied by the (non-square) matrix converting the 150 outputs to all outputs including the derivatives expressions. In this layer, all coefficients are known.

All these layers are followed by a ReLU activation function, apart from the maxpooling layer
which is not followed by any, and the last dense linear layer, which is coupled to an
activation function of type hyperbolic tangent.

In total our CNN comprises 687,815 free model parameters and is implemented
using PyTorch. Our sample of spectra is split into a 80\% training set, a
10\% validation set and a 10\% test set. We also experimented with different
splits, but the results remain the same. The optimization is carried out
via the NAdam algorithm, employing an epoch-dependent learning rate: $10^{-3}$
for the first 50 epochs, $10^{-4}$ for the subsequent 50 epochs, and $10^{-5}$
for the remaining 250 epochs. We use the L1 norm as our
loss function with a sum reduction type. We find that our CNN model is straightforward
to train and produces accurate results. Our CNN performances are attested by
several metric scores applied on the validation set. With the inclusion of
derivative expressions, the average \(R^2\) score is $0.84$ where the average
is taken across all resulting spectral point plus derivatives, the mean squared error
(MSE) is $0.0004$, the mean absolute error (MAE) is $0.0027$ and the root mean
squared error (RMSE) is $0.004$. In contrast, omitting the derivatives from the
final score yields an average $R^2$ score of $0.9960$, an MSE of $9.4374 \times 10^{-6}$,
a MAE of $0.0013$, and an RMSE of $9.4374 \times 10^{-6}$, all of which attest
to the excellent performance of our CNN.

\begin{figure*}
    \centering
    \includegraphics[width=0.95\textwidth]{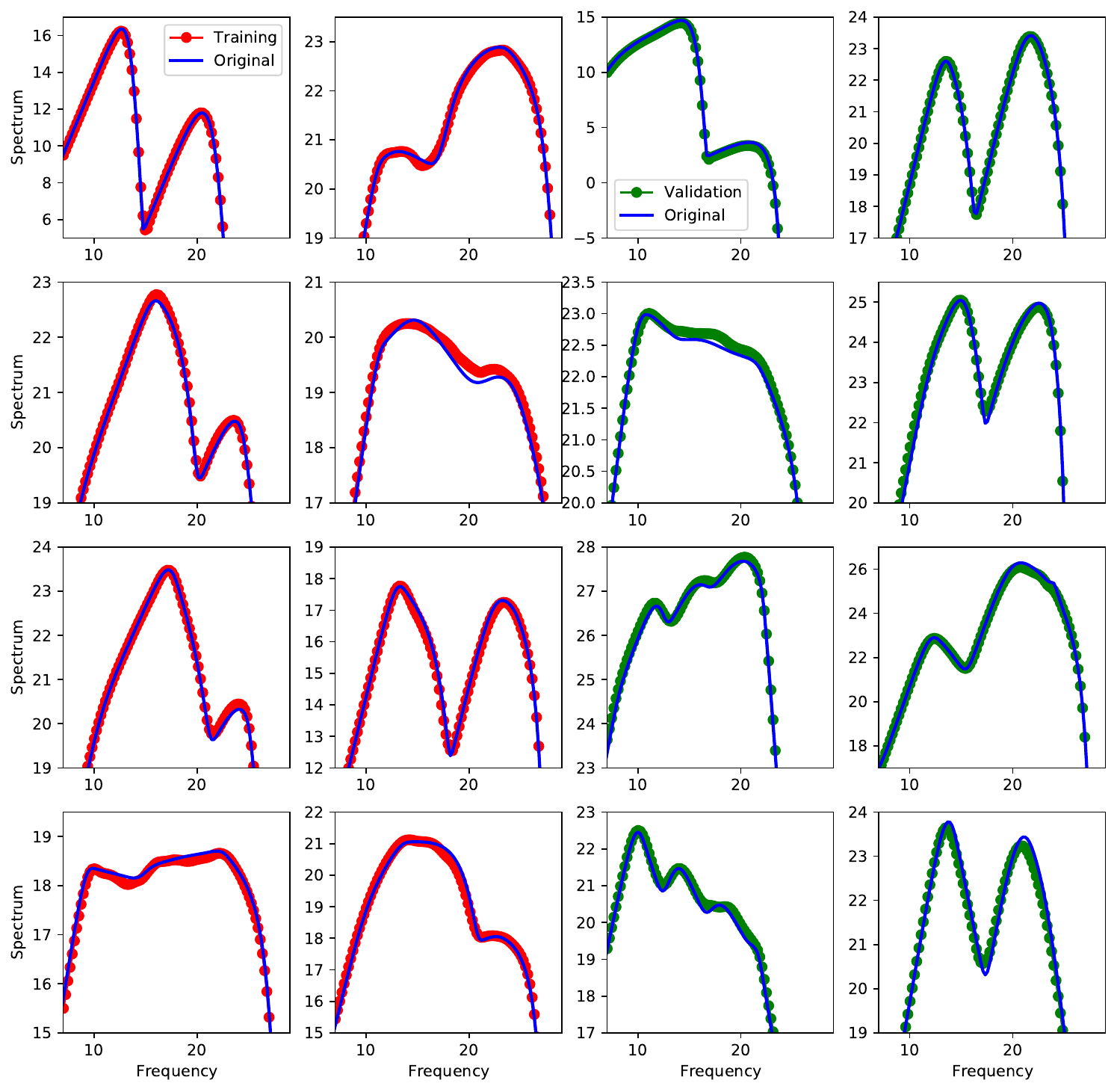}
    \caption{ Comparison between the $\nu F_\nu$ spectra computed by the CNN (dotted line)
    and by {\it SOPRANO} (solid line) before unit conversion and expression in the observer
    frame. The x axis is the comoving frequency in unit of Hz. \textit{Left Panel:} Spectra
    from the training set. \textit{Right Panel:} Analogous spectra from the validation
    set. This figure shows the large diversity of spectra the CNN must be (and is) able
    to reproduce, the accuracy at which it reproduces them and the wide spam of the
    typical emitted power across the leptonic SSC model.}
    \label{fig:CNN_spectra}
\end{figure*}

In Fig. \ref{fig:CNN_spectra}, the two leftmost columns display representative
examples of $\nu F_{\nu}$ spectra from the training set. They are superposed with their
corresponding spectra as computed by the CNN. In contrast, the two
rightmost columns of Fig. \ref{fig:CNN_spectra} feature spectra from the
validation set, that is to say that were not used to train the
CNN. These are also compared with their respective spectra generated by our CNN
for comparative analysis. Despite a wide spread in normalization, the agreement
between the original {\it SOPRANO} spectra and their corresponding CNN-generated spectra
is remarkably high, spanning multiple orders of magnitude in both power and frequency.
Notably, key features such as the synchrotron peak and the inverse Compton peak
are accurately reproduced, once more attesting to the accuracy of the CNN model
in reproducing the complex spectra produced from {\it SOPRANO}.

We note that the accuracy for some spectra is lower than for others. For instance, the second and third spectra on the second line is slightly off around frequency $10^{20}$Hz. We find that this happens at the boundary of the parameter space, as there is less information for the model to be train. On the other hand, these parameters are not expected to be relevant for the analysis of blazar SED, but have to be included to form regular continuous and independent parameter distributions.

\begin{figure*}
    \centering
    \includegraphics[width=0.48\textwidth]{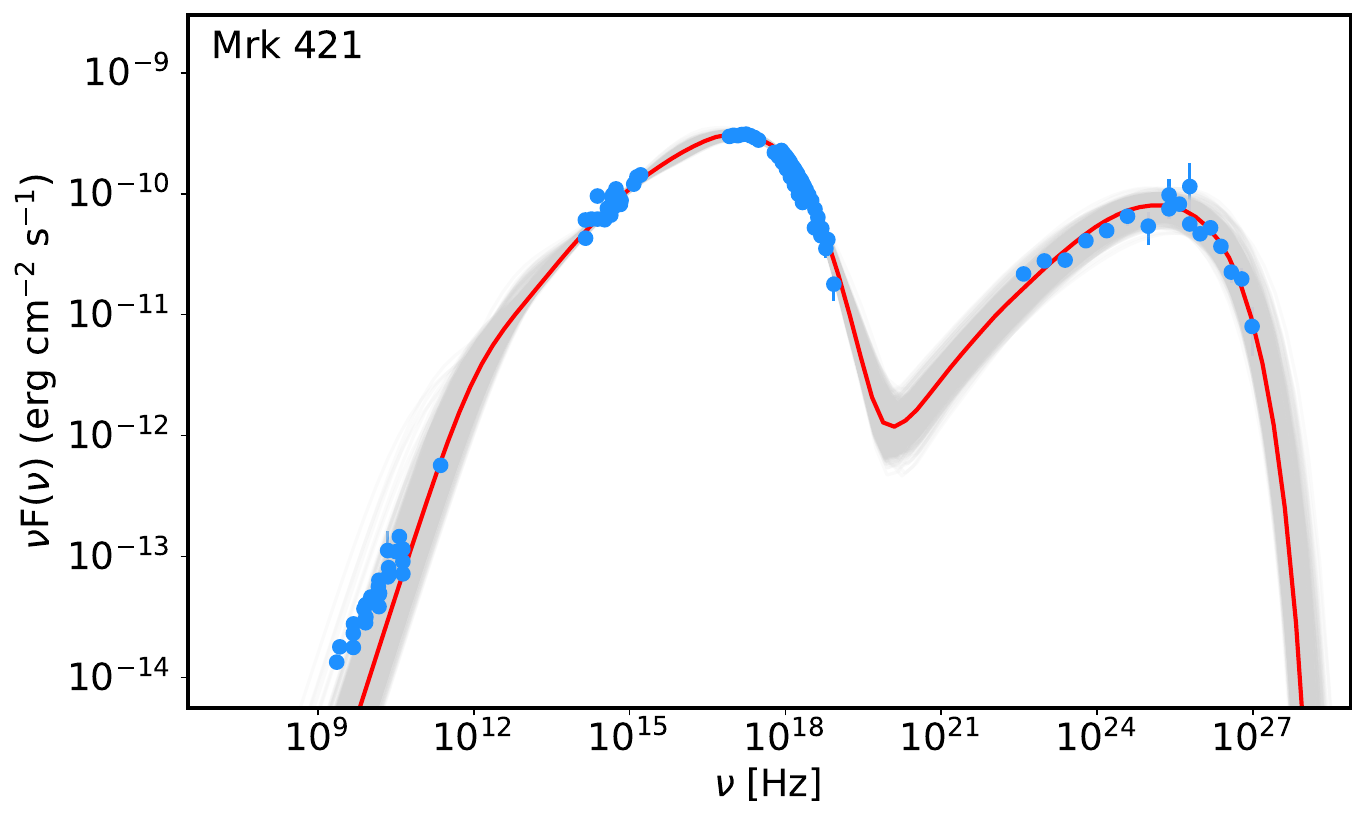}
    \includegraphics[width=0.48\textwidth]{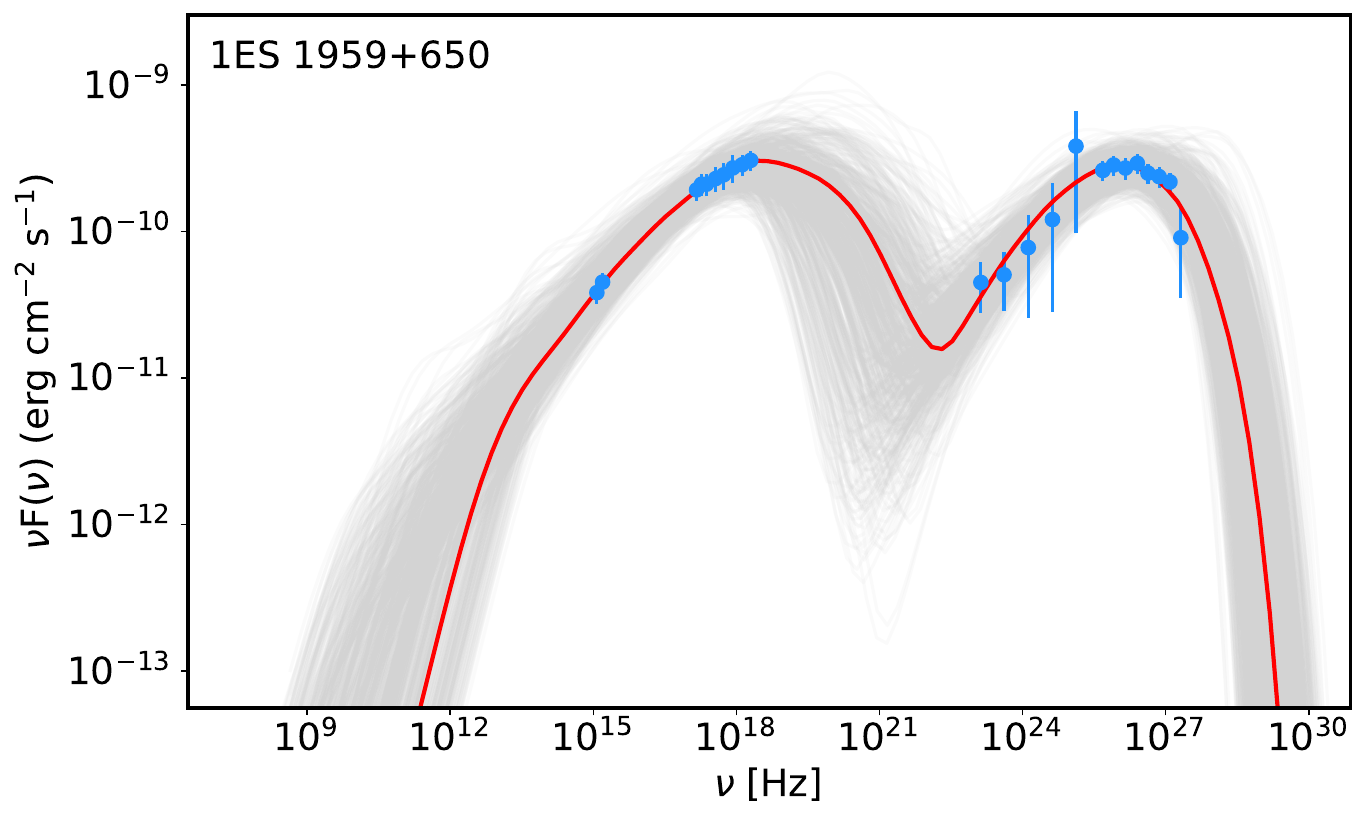}
    \caption{The broadband SEDs of Mrk 421 during the 4.5-month-long multi-wavelength campaign in 2009 (left) and of 1ES 1959+650 on the 14$^{\rm th}$ of June 2016 (right). The data and its errors are in blue, the red line is the model corresponding to the best parameters, i.e., maximizing the likelihood, and the grey spectra represent one in 10 randomly selected samples from the MCM sampling, representing the model uncertainty. The data are corrected for EBL absorption.}
    \label{sed}
\end{figure*}

\begin{table}
    \centering
\caption{Parameters describing the SEDs in Fig. \ref{sed} and \ref{sed:1ES}. The two middle columns correspond to models with all 7 parameters free, while in the rightmost column, $R$ and $\delta$ are linked through the variability time.}
\label{tab:param}
    \begin{tabular}{c||c||c|c}
    \hline
    Parameters  &  Mrk 421 &  1ES 1959+650  & 1ES 1959+650 \\\hline 
    \( p \) & \( 2.16 \pm 0.08 \) & \( 2.26 \pm 0.16 \) & \(  2.12 \pm 0.14 \) \\
    \( \log_{10}(\gamma_{\rm max}) \) & \( 4.87 \pm 0.16 \) & \( 6.41 \pm 0.45 \) & \(  5.85 \pm 0.48 \) \\
    \( \log_{10}(\gamma_{\rm min}) \) & \( 2.49 \pm 0.22 \) & \( 3.20 \pm 0.69 \) & \( 1.56 \pm 0.68 \) \\
    \( \delta \) & \( 31.01 \pm 8.08 \) & \( 22.49 \pm 9.54 \) & \( 40.03 \pm 5.69 \) \\
    \( \log_{10}(B/[\rm G]) \) & \( -1.17 \pm 0.28 \) & \( -1.62 \pm 0.38 \) & \( -0.72 \pm 0.23 \) \\
    \( \log_{10}(R/[\rm cm]) \) & \( 16.43 \pm 0.44 \) & \( 16.82 \pm 0.38 \) & \( 15.44 \) \\
    ~~~\( \log_{10}(L_{\rm e}/[\rm erg\:s^{-1}]) \)~~~ & ~~~ \( 43.32 \pm 0.18 \) ~~~ & ~~~ \( 43.71 \pm 0.32 \) ~~~ & ~~~ \( 43.41 \pm 0.19 \) ~~~ \\\hline
    \(\log_{10}(L_{\rm B}/[\rm erg\:s^{-1}]) \) & \( 43.08 \) & \( 42.68 \) & \( 42.13 \) \\
    \hline
      &  All parameters free & All parameters free & Variability time constraint
  \end{tabular}
  \label{params}
\end{table}

\section{Modeling the broadband SEDs of Mrk 421 and 1ES 1959+650}

\label{sec:example}

To demonstrate the efficieny of
our approach based on CNN in fitting and interpreting the SEDs
of blazars, we model in this section the
observed broadband dataset of two well-studied
sources namely, Markarian 421 (Mrk 421) and 1ES 1959+650.
Our analysis assumes uniform priors for the electron index $p$
and the Doppler boost $\delta$, and log-uniform priors for all remaining
parameters, namely $R$, $B$, $L_e$, $\gamma_{\rm min}$, $\gamma_{\rm max}$.
We assume a Gaussian likelihood and sample the posterior distributions with MultiNest \citep{FHB09}, a nested sampling algorithm designed for efficient Bayesian inference. We assume 1000 active points and a tolerance of $0.5$ to ensure efficient sampling and convergence. MultiNest offers a number of advantages, including computational efficiency and the ability to robustly handle multi-modal posterior distributions, which is a distinct possibility given the high dimensionality and complexity of the parameter space.

\subsection{ Markarian 421}
Located at a redshift of $z = 0.031$, Mrk 421 is one of the most extensively
monitored blazars as it is the brightest blazar in the
extragalactic X-ray sky. Owing to its proximity and brightness, the broadband
emission features of Mrk 421 have been thoroughly investigated at all wavelengths from radio to VHE \grays. In 2009, a
4.5-month-long multi-wavelength campaign was conducted,
yielding an unprecedented volume of simultaneous data \citep[][]{2011ApJ...736..131A}.
The observed SED is presented in the left panel of Figure \ref{sed}, where
the set of data set is obtained from \citet{2011ApJ...736..131A}. 
We performed a fit to the SED, excluding data below $10^{11}$ Hz, as emission in the radio band can be self-absorbed, implying that it is dominated by the outer regions of the jet. The best-fit parameters are listed in the left column of Table \ref{tab:param}. The left panel of Figure \ref{sed} displays the model uncertainty in grey and the best model, based on the best-fit parameters, in red. The posterior distribution functions are provided in Figure \ref{fig:posterior_Mrk421} in the appendix.

The model displayed in the left panel of Figure \ref{sed} accurately reproduces the observed data above 225 GHz. Given the current parameter set, self-absorption dominates below \(1.3 \times 10^{13} \, \text{Hz}\), making it impossible to model lower frequency data. The parameters we obtained are somewhat in agreement with the values determined by \cite{2011ApJ...736..131A}, who used a three-component power-law function to fit the broadband SED. In their model, the electron distribution between $\gamma_{\rm min}=8.0\times10^2$ and $\gamma_{\rm brk,1}=5.0\times10^4$ has an index of $2.2$, which is consistent with our estimated value of $p=2.16$ (for the errors see Table \ref{params}). In our approach the main difference is that we achieve an acceptable fit by assuming a single electron index for the injection, which is consistently evolved under the influence of radiation cooling. In our case, the synchrotron cooling  would affect the spectrum at a frequency of $\approx5.6\times10^{17}$ Hz. This is above the maximum frequency defined by  $\gamma_{\rm max}=7.37\times10^4$ ($8.5\times10^{16}$ Hz). Consequently, an electron spectrum with a power-law index of $p=2.16$?,  above \(\gamma_{\rm min}=2.10\times10^2\) is sufficient to reproduce the observed spectrum.
Our fit indicates that the magnetic field is around $B=6.70\times10^{-2}$ G, which is in agreement with the value from \cite{2011ApJ...736..131A}  within the uncertainties. The dissipation radius we obtained, $R = 2.96\times10^{16}$ cm, is somewhat close to the value estimated in their model which was derived based on the variability time. We further find that the total luminosity of the electrons, $(L_{\rm e} = 2.09\times10^{43} \, {\rm erg \, s^{-1}}$, is of the same order of magnitude as the magnetic field luminosity $L_{\rm B} = 1.19\times10^{43} \, {\rm erg \, s^{-1}}$, calculated as $L_{B}=\pi c R^2 \delta^2 B^2/8\:\pi$. This suggests that the system is close to equipartition.

\subsection{1ES 1959+650}

Blazar 1ES 1959+650, at $z=0.048$, is another bright
blazar known for frequent flaring  across the optical, X-ray, and TeV bands.
The X-ray and \gray\ (TeV) flares often occur simultaneously, although orphan \gray\ flares
have also been observed. This suggest that the same population
of electrons is responsible for emissions in both bands. The source was in an active state
from April to November 2016, during which the MAGIC telescopes observed major VHE \gray\
flares on June 13 and 14, as well as July 1, 2016 \citep{2020A&A...638A..14M}.
The multi-wavelength campaigns conducted during these flaring periods also enabled the
accumulation of data across lower-frequency bands,
providing a comprehensive view of the flaring activities. In this study, we focus on
modeling the flare observed on the 13$^{\rm th}$ of June 2016. We retrieved the data of
the flare from \cite{2020A&A...638A..14M}. We note that the data are 
corrected for extragalactic background light (EBL) absorption. If it was not the case, our
numerical model includes the possibility to incorporate EBL absorption, via the model of \citet{2011MNRAS.410.2556D}. 

The fit to the data obtained during the flaring
activity of 1ES 1959+650 is depicted in the right panel of Figure \ref{sed},
and the corresponding parameter posterior distributions are provided in
Figure \ref{fig:posterior_1ES1959}. The best-fit parameters are summarized
in Table \ref{tab:param}. The data suggest that the synchrotron peak should
occur at frequencies $\gtrsim 10^{20}$ Hz, enabling the X-ray data to
constrain the power-law index of the electron injection function at $p=2.26$. In contrast
to the case of Mrk 421 where the X-ray data define the
high-energy tail of the synchrotron component, the value
of the parameter $\gamma_{\rm cut}=2.57\times10^{6}$ is
not well-constrained in this case. It is determined solely
by the last data of the MAGIC spectrum, which
have a large uncertainty. The interpretation of
the parameter is also difficult because of the EBL effect at these
high frequencies. The fit to our model constrains the magnetic field to be $2.4\times10^{-2}$G and the Doppler boost
$\delta$ to be $22.49$. The parameters $p$, $\gamma_{\rm cut}$, and $\delta$ are similar to those proposed by
\cite{2020A&A...638A..14M}, but the magnetic field and the radius $R$ differ significantly.

The dissipation radius $R = 6.69\times10^{16}$cm is rather large and the
value of the Doppler factor is average, $\delta = 22.49$, which leads to an estimated variability time of $t_{\rm var} \sim 10^5$s, which is much longer than the reported variability time of approximately
36 min \citep{2020A&A...638A..14M}. Although our fitting procedure generally treats the radius $R$ and $\delta$ as independent variables, we can easily couple these parameters by specifying the variability timescale and removing one of them from the model parameter. To illustrate this approach, we set the radius to be $R = c \delta t_{\rm var}$ and retain $\delta$ as a free parameter.
In order to not jump outside of the parameter range, the bounds on $\delta$ are changed to
$\delta_{\rm min} = {\rm max}(3, R_{\rm min}/(c t_{\rm var}))$ and
$\delta_{\rm max} = {\rm min}(50, R_{\rm max}/(c t_{\rm var}))$.

\begin{figure}
    \centering
    \includegraphics[width=0.48\textwidth]{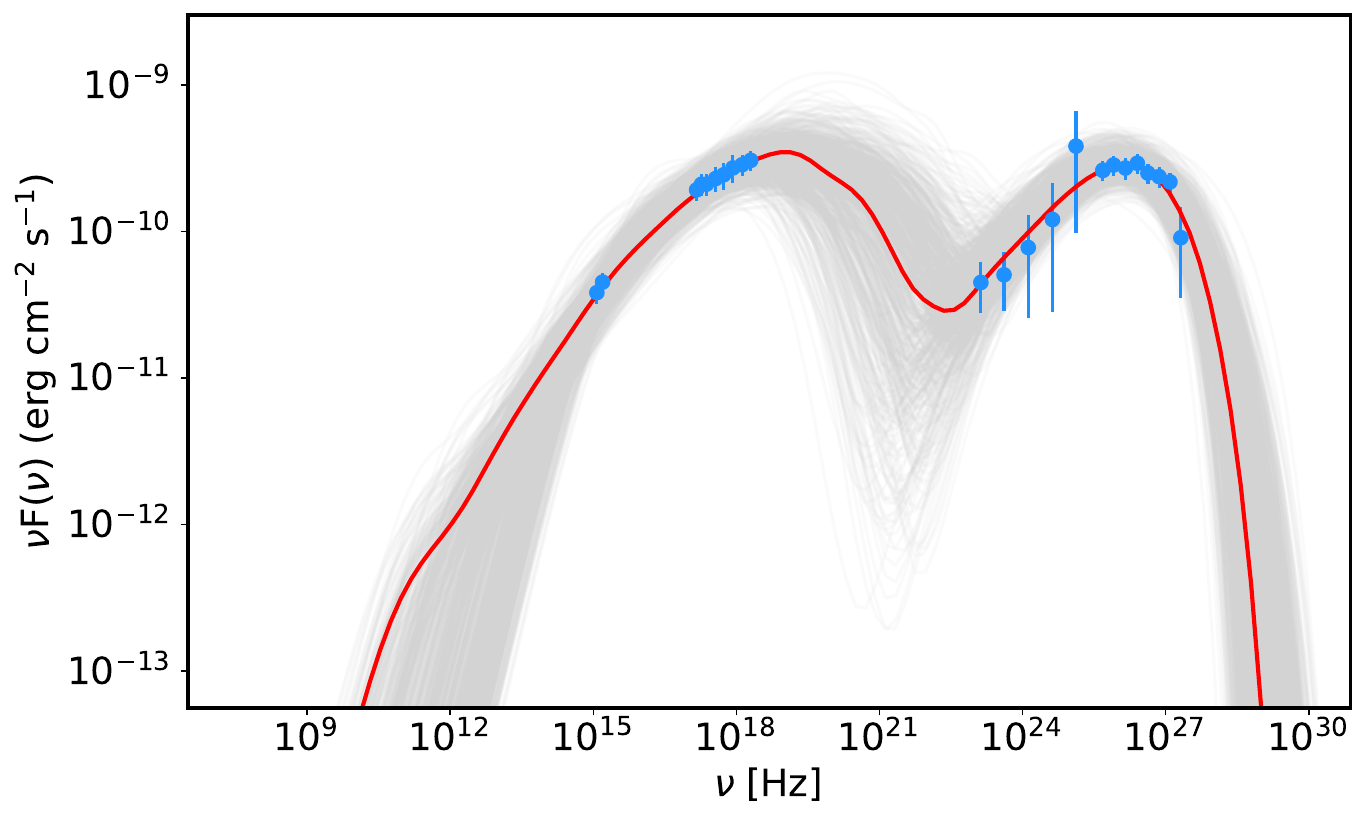}
    \caption{The same SED of 1ES 1959+650 as presented in Fig. \ref{sed}, but fitted with a model where $R$ is constrained by a variability of 35 minutes.} 
    \label{sed:1ES}
\end{figure}

The fit results are illustrated in Figure \ref{sed:1ES}, while the parameter posterior distributions are presented in Figure \ref{fig:posterior_1ES1959_tvar}. The best-fit parameters are listed in the rightmost column of Table \ref{tab:param}. A
significant difference is observed in the value of the Doppler boost parameter,
$\delta$, which has shifted to larger values, compared to
$\delta \approx 22.49$ in the previous scenario. This indicates that the compact emission region is moving at a higher velocity. Additionally, the magnetic field density in this case is larger — $B \sim 0.24$G as opposed to $B\sim 0.02$G in the previous case — which influences the electron cooling process. In the first case, synchrotron cooling is inefficient for all electrons.
However, in the second case, the synchrotron cooling is efficient for the highest energy electrons, and a cooling break occurs at $\sim 2 \times 10^{17}$ Hz, resulting in the X-ray emission to be produced by
cooled electrons.

\section{Conclusion}\label{sec:conc}

In this work, we presented a new approach to fit multi-wavelength SEDs of blazars with numerically intensive models. Indeed, there is a clear gap between the computational resources needed for each model evaluation and the analysis,
fitting, and detailed interpretation of multi-wavelength (and soon, multi-messenger) data for blazars. To bridge these two aspects of blazar SEDs analysis, we developed a neural network that
can be trained on different computationally demanding numerical models. In this study, the CNN is trained on a large set of SSC spectra generated
by {\it SOPRANO}, taking into account all relevant
cooling processes and the pair creation process. Our surrogate model achieves high accuracy, is computed in a relatively short time $\sim$ms, includes the self-consistent cooling of the electron, and enables on-the-fly fits to data. We demonstrate the performance of the CNN by fitting the multi-wavelength observations of two BL Lac objects, namely Mrk 421 and 1ES 1959+650, thereby constraining the parameters of the SSC model and obtaining their posterior distributions.

The significant advantage of the method proposed in this work is its computational speed; the model performs fast independently of of the considered physical processes and is expected to do so when hadronic processes will be included. However, a key limitation of this approach is the initial requirement for the substantial computational resources to generate the spectra needed for training the CNN. Once this initial step is completed,
our methodology enables efficient and straightforward analysis of blazar SEDs.
The low computational cost of the model evaluation via the CNN offers the advantage of enabling more sophisticated data fitting techniques. In
future works, this efficiency will permit us to allocate computational resources
for model forward-folding. Specifically, instead of using pre-analyzed data, we plan to utilize raw observational data in conjunction with the response functions of various instruments, such as Swift-XRT and Fermi-LAT.
This integration will be facilitated through the use of 3ML \citep{VLY15}, a
framework specifically designed to combine analyses from different instruments
across energy bands into a unified, coherent picture.

In this study, we trained the first convolutional neural network to accurately model the radiative signatures associated with the SSC model. This approach provides
a novel framework for fitting the SED of blazars, and we
intend to further apply it to other models of
blazar SEDs. Specifically, we plan to implement additional
computationally intensive models based on external Compton and hadronic scenarios,
for which the CNN will be trained. This set of models will facilitate the
interpretation of a large variety of blazar SEDs, spanning various wavelengths,
time periods, and sources.

We believe that the approach outlined in this paper has the potential to provide significant advances of our understanding of blazars by enabling the fitting of self-consistent models to their SEDs. To facilitate broader analysis and interpretation, the model developed here will be made publicly available on the Markarian Multiwavelength Datacenter (mmdc)\footnote{\url{http://www.mmdc.am}}. Users will be able to interact with an interface to reproduce single snapshot SEDs by specifying model parameters. Additionally, users will be able to
perform fits after uploading their data (if necessary), which will provide them with the parameters that best describe the observed data, along with their posterior distributions. It should be noted that, as of the current time, this will be the only public tool available for performing fits with self-consistent model of blazar SEDs.

Not only the CNN and the associated methodology could be applied to several model
of blazar as demonstrated here, but we believe that it is sufficiently general
and robust to also be used in spectral and temporal analysis of gamma-ray bursts
prompt and afterglow phase, multi-wavelength temporal evolution of kilonovae
\citep[e.g.][]{BvL23}, and for the spectral interpretation of X-ray binaries.

In summary, this study represents a pioneering effort in employing convolutional neural networks for the efficient and accurate modeling of blazar SEDs. We have introduced a flexible and efficient methodology for self-consistent blazar modeling, which holds the potential for deepening our understanding of blazar physics. With the tool made publicly available through the Markarian Multiwavelength Data center, researchers will be able to perform state-of-the-art, self-consistent analyses of multi-wavelength—and soon, multi-messenger-data from blazar observations.

\section*{Acknowledgements}
DB, HD and AP acknowledge support from the European Research Council via the
ERC consolidating grant $\sharp$773062 (acronym O.M.J.). NS, SG and MK
acknowledge the support by the Higher Education and Science Committee of
the Republic of Armenia, in the frames of the research project No 23LCG-1C004.

\section*{Data availability}
All the observational data used in this paper is public. The convolutional neural network used
to fit the SEDs can be shared on a reasonable request to the corresponding author. In addition, it is publicly available through the Markarian Multiwavelength
Datacenter (\url{http://www.mmdc.am}).

\bibliographystyle{mnras}
\bibliography{biblio}

\appendix

\section{Parameter posterior for Mrk 421 and 1ES 1959+650}

\begin{figure*}
    \centering
    \includegraphics[width=0.95\textwidth]{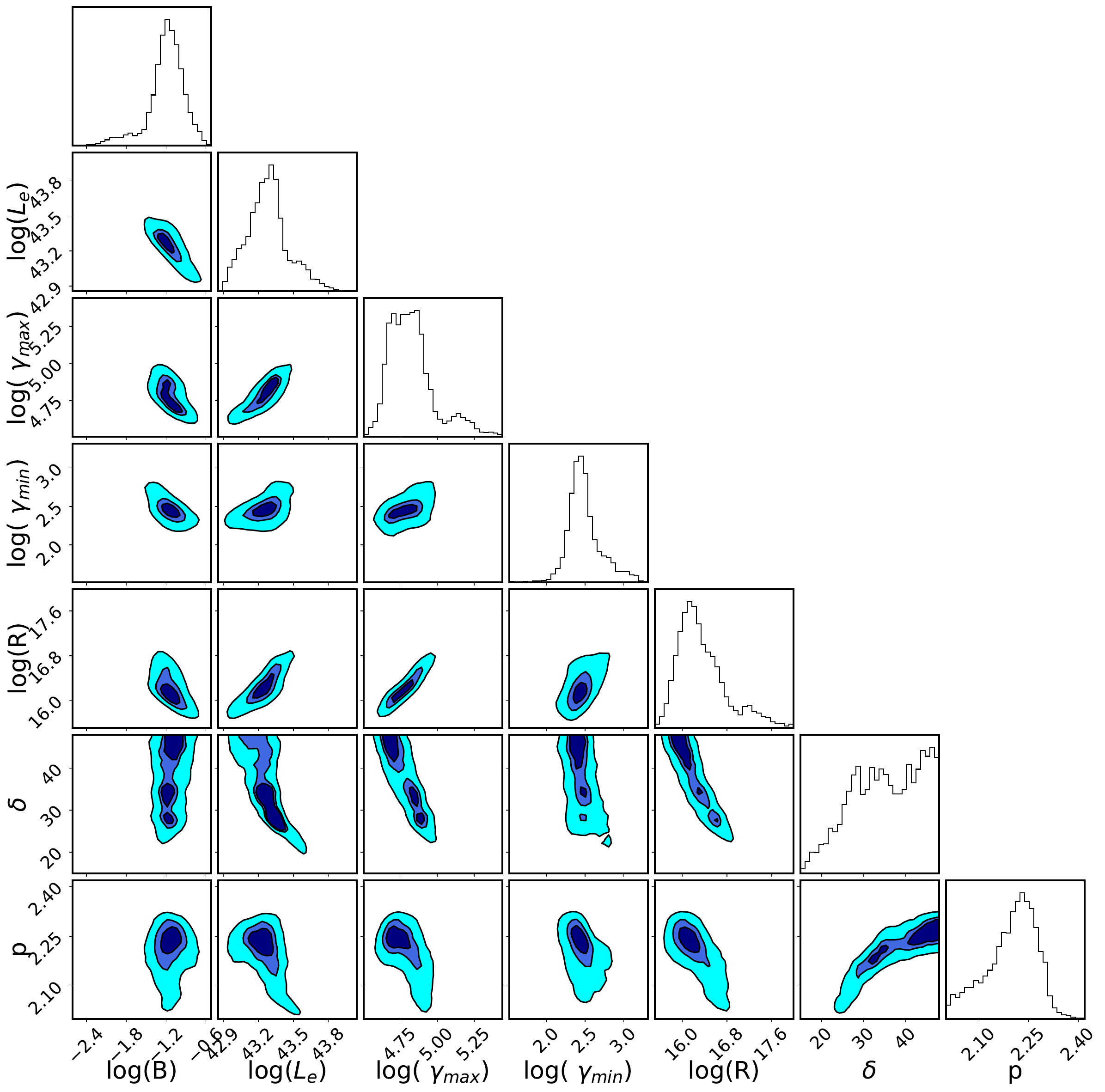}
    \caption{Parameter posterior distributions for Mrk 421 during the multi-wavelength campaign of 2009. The contours gives from outward to inward the 20\%, 40\%, and 75\% confidence regions. Apart from the radius $R$, all parameters are well constrained.}
    \label{fig:posterior_Mrk421}
\end{figure*}

\begin{figure*}
    \centering
    \includegraphics[width=0.95\textwidth]{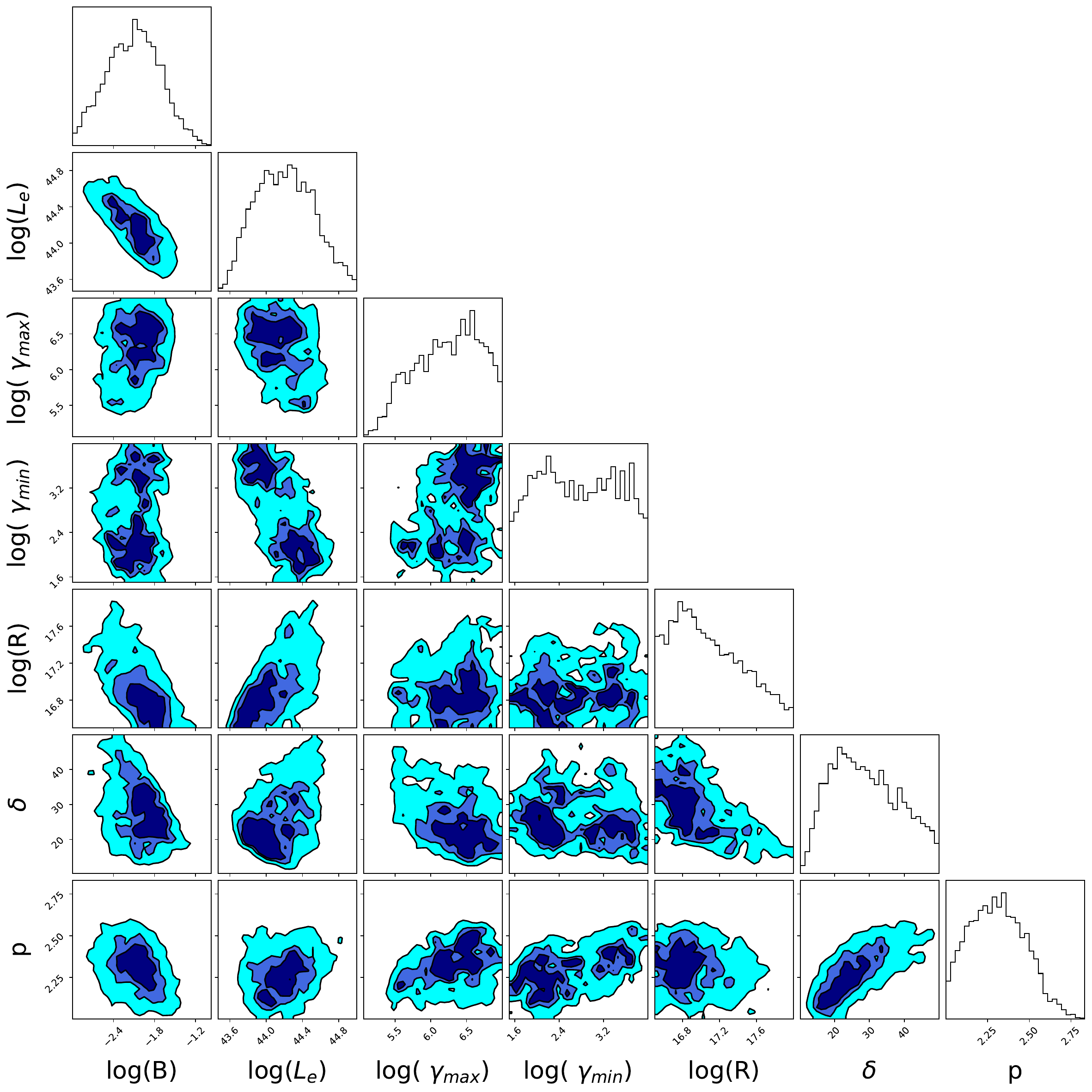}
    \caption{Parameter posterior distributions for 1ES 1959+650 without imposing a variability constraint. The magnetic field, the electron luminosity and the electron index are well constrained. In contrast, the other parameters remain somewhat unconstrained due to the high uncertainty in the position of peak energy of the synchrotron bump.
    }
    \label{fig:posterior_1ES1959}
\end{figure*}

\begin{figure*}
    \centering
    \includegraphics[width=0.95\textwidth]{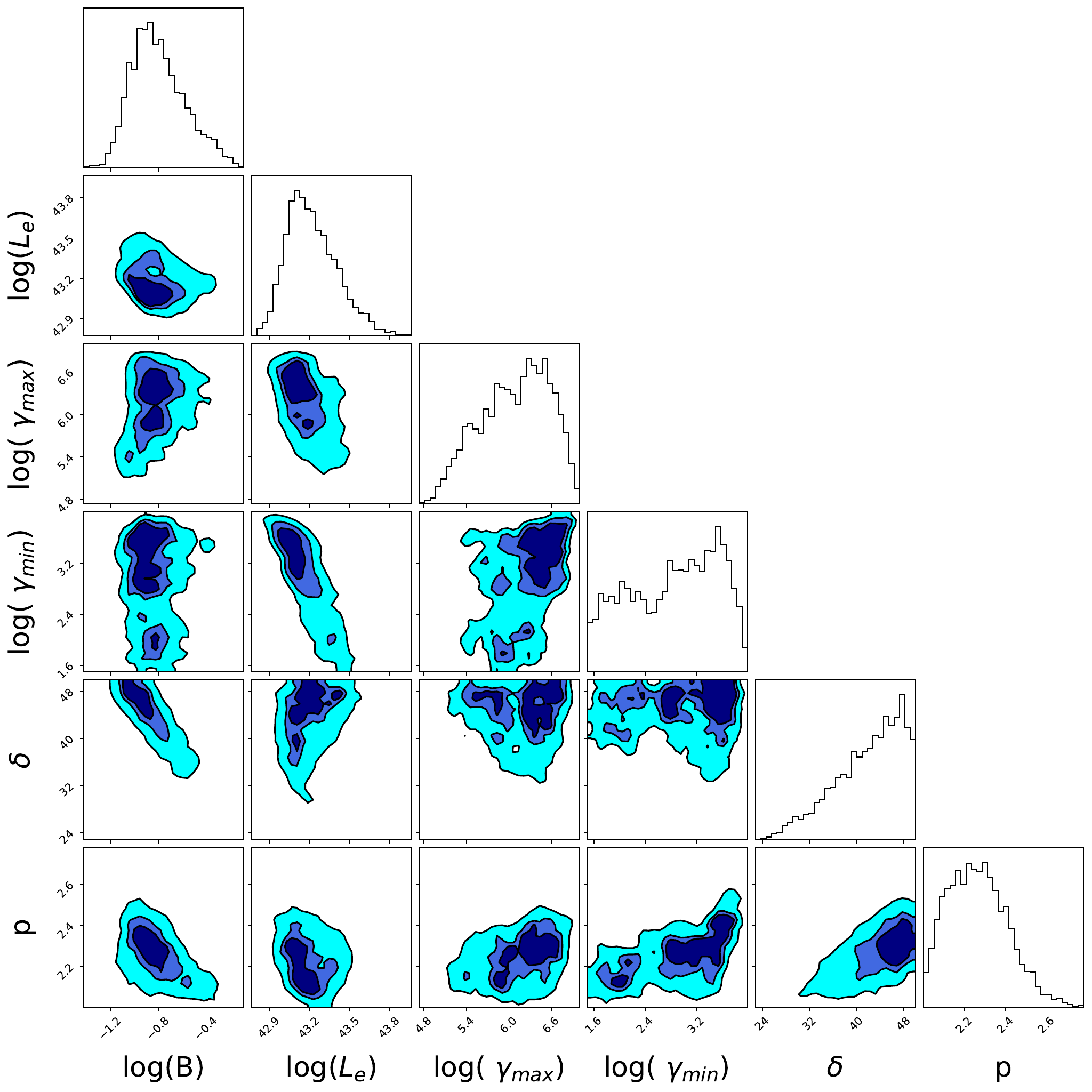}
    \caption{Parameter posterior distributions for 1ES 1959+650 assuming a variability time of 35 minutes to link the radius and the Doppler boost. Additionally, there exists a large uncertainty on the minimum electron Lorentz factor $\gamma_{\rm min}$}.
    \label{fig:posterior_1ES1959_tvar}
\end{figure*}

\end{document}